\documentclass{article}
\usepackage{bigsky2007}
\usepackage{graphicx}

\frompage{000} \topage{000}                                              
\def\rightmark{Multi-strange baryon correlations at RHIC}
\def\leftmark{Betty I. Abelev (for the STAR Collaboration)}

\title{Multi-strange Baryon Correlations at RHIC}
\authors{
{Betty I. Abelev (for the STAR Collaboration)%
}\\[2.812mm]
{\normalsize
\hspace*{-8pt} University of Illinois at Chicago,
Chicago, IL\\[0.2ex]
}}

\abstract{Multi-strange baryon azimuthal correlations were observed
in $d$+Au and Au+Au data taken by the STAR detector at RHIC.  We
extract the same-side per-trigger yields for a variety of strange
particle species and trigger \pt,\ but do not observe any species
dependence. We also report the observation of an elongation in
    the \de\ direction of the $\Xi$ correlation peak, the ridge. Comparing the
    same-side yields in \dau\ and \auau\ data, we conclude that the same-side yield
    of the azimuthal-only $\Xi$ correlations is greatly enhanced by the ridge yield.
    We also report the first observation of a strong same-side peak in the \Om\ baryon
    triggered azimuthal correlations.  }

\newcommand{\Om}{\ensuremath{\Omega}}

\newcommand{\La}{\ensuremath{\Lambda}}
\newcommand{\pt}{\ensuremath{p_{\perp}}}

\newcommand{\df}{\ensuremath{\Delta\phi}}
\newcommand{\de}{\ensuremath{\Delta\eta}}
\newcommand{\dau}{\ensuremath{d}+Au}
\newcommand{\auau}{Au+Au}

\begin{document}

\maketitle
\setcounter{page}{1}

\section{Introduction}\label{intro}
In the heavy-ion collisions produced at Relativistic Heavy Ion
Collider (RHIC) in Brookhaven National Laboratory, we observe a
formation of a very dense, hot medium. The formation of this medium
is manifested, among other effects, by copious production of strange
quarks and anti-quarks.

An enhancement of strange particle production has long been
predicted as a signature of a large thermalized system with partonic
degrees of freedom \cite{KochMullRaf}. The argument runs two-fold.
Firstly, the threshold for strangeness production in a partonic
medium is the mass of a strange quark-antiquark pair ($\sim$200
MeV).  This is less than half the energy threshold of strangeness
production in a hadron gas.  Secondly, due to the comparatively
large volume of the central \auau\ interaction region, the system
can be described by a grand-canonical ensemble, which makes the
production of strange particles relatively easier the more
strangeness they contain. Thus, we would expect a proportionally
larger number of \Om\ (sss) and $\Xi$ (ssd) baryons in most central
\auau\ collisions than, for example, in \dau.\ Indeed, this has been
observed \cite{matt}. However, there is much to learn about
particle-production mechanisms in this medium. One of the puzzles to
come out of RHIC is the enhancement of the baryon to meson ratio at
intermediate \pt\ ($2< $\pt\ $<6$ GeV/c) in the most head-on \auau\
collisions with respect to the ratio observed in $p+p$ and the most
peripheral \auau\ collisions \cite{matt}. This enhancement seems to
be qualitatively explained by coalescence-recombination models
\cite{matt}. A typical recombination model calculates the
contributions of the soft and hard parton production to a given
particle spectra.  One of the recombination models, proposed by Hwa
et al., \cite{bib1} predicts that the vast majority of \Om\ baryons
produced in the intermediate \pt\ region are from recombination and
not from fragmentation.  The authors also suggest that we can test
the extent of the soft production of the s-quark by using the
azimuthal correlations with \Om\ triggers. Azimuthal correlations
allow us to study jet production statistically, without full
reconstruction of the jet energy. The back-to-back jets manifest
themselves as Gaussian peaks, separated by $\pi$ radians in azimuth.
In the Hwa model no such peaks would be observable due to the
overwhelming majority of the \Om\ triggers emanating from ``soft"
sources.

In this work we examine this assumption and present an azimuthal
correlation with a non-negligible \Om\ triggered same-side peak.  We
also present $\Xi$ baryon correlations, where the statistics are
more abundant and thus a closer examination of a result is possible.

\section{Data and the experimental setup}\label{exp}

    \subsection{Data}
    The data presented in this study were obtained in the third (\dau) and fourth (\auau) year of
    RHIC operations, using the STAR detector \cite{bib3}.
    Both data sets were taken at $\sqrt{s_{NN}}=200$ GeV. Whereas the 20 million \dau\ events
    used in the analysis came from a minimum bias data sample, the 24 million \auau\ events
    presented were obtained using the STAR central trigger.  This trigger setup uses the number
    of forward neutrons that remain after the collision to estimate the impact parameter and
    thus the collision's centrality.  The fraction of the total cross-section that corresponds
    to this trigger is 0-12\% most central (or most violent, i.e., events with the smallest
    impact parameter) events.  However, only the events that matched the charged track multiplicities
    of the minimum bias 0-10\% sample were used.  In the \dau\ data, to avoid pile-up events, the STAR Central Trigger
    Barrel (CTB) was used ensuring that all particles used in a correlation came from the
    same event (all particles had to pass through the CTB at the same time). In the \auau\ data, the accepted event
    vertex was restricted to $\pm25$ cm from the center of the detector.  For the \dau\ data set,
    all usable events were within 50 cm of the detector center.

    \subsection{Tracking and particle reconstruction}
    The main component of the STAR detector is its large acceptance Time Projection Chamber (TPC) \cite{bib4},
    which covers 2$\pi$ radians in azimuth and 3.6 units of pseudo-rapidity.  The TPC is situated inside
    a 0.5 T solenoidal magnetic field, which allows for the determination of charged particle momenta starting
    from transverse momenta of 100 MeV/c and identification of kaons, pions and protons using their energy
    loss (dE/dx) in the TPC gas.

    The identification of strange baryons is done using the geometries of their most abundant channel decay
    topologies, $\Xi^{\pm}\rightarrow\Lambda^0+\pi^{\pm}$ (99.9\%) and $\Omega^{\pm}\rightarrow\Lambda^0+K^{\pm}$
    (68\%).  First, the $\Lambda$ is identified via its characteristic V-shaped decay.  Then, using the dE/dx
    identified pions or kaons, each multi-strange baryon is reconstructed by selecting meson-$\Lambda$ pairs
    that fit our selection criteria of a secondary cascade vertex.


\section{Analysis method}\label{details}

\subsection{Constructing a correlation}
The correlation function is composed of particle pairs, made up of
{\it trigger} and {\it associated} particles. In the study
presented, the trigger particle is the multi-strange baryon, and the
associated particle is any primary (i.e., originating at the
collision vertex) charged hadron from the same event that is not a
decay daughter of the trigger particle. Since the trigger hadron is
assumed to be the leading particle in a jet, its \pt\ is always
higher than the \pt\ of the associated particle.

The correlation is constructed as follows: using the geometry of a
multi-strange baryon decay, we calculate the invariant mass of our
candidates that fall in the pre-specified \pt\ range, and select
only the ones that fall under the mass peak (the actual resolution
depends on the particle and the data set). The candidates must
originate at the collision vertex.  After we select the trigger
candidate, the same event is scanned for an associated track with
\pt\ greater than a pre-set threshold (1.5 GeV/c in all cases,
except 2-dimensional correlations), but less than the \pt\ of the
trigger.  Then we calculate the azimuthal angle of each at the
collision vertex, and then calculate the difference (\df) between
the two. Similarly, we calculate the \de\ of the two particles. We
correct each correlation by the efficiency of finding a charged
track at a given \pt.\ In \dau\ data the efficiency is a uniform
89\% for \pt\ $>$1.5 GeV/c. For \auau\ data the efficiency varied
from 72 to 78\% depending on the centrality of the event and the
\pt\ of the charged track.  We also correct the entire correlation
function for detector acceptance in both azimuth and
pseudo-rapidity. This is done by first constructing a correlation
function by combining random trigger and associated particles from
separate pools collected over all events. The correction function is
then normalized to its maximum value and the true correlation
function is then divided by this constructed correction function.

\subsection{Background subtraction: Flow determination}\label{sect:flow}

In Au+Au collisions, the underlying background shape is a direct
consequence of the azimuthal anisotropy of the collision and of the
underlying partonic flow. The 2$^{nd}$ harmonic of the elliptic flow
is sinusoidal in shape and can be described by expression
\ref{eq:v2}:

\begin{equation}
B\times 2v_2^{assoc}v_2^{trigg}\cos{2\Delta\phi}
\label{eq:v2}
\end{equation}

where $B$ is the height of the uncorrelated background, and
$v_2^{assoc}$ and $v_2^{trigg}$ are the second harmonic coefficients
of the Fourier expansion of the elliptic flow for trigger and
associated particles respectfully.   In STAR, there are several
distinct methods for elliptic flow determination.  The most commonly
used are the 4-particle cumulant ($v_2\{4\}$), and the so-called
``Reaction Plane" methods \cite{bib5}.  Due to the small
eccentricity of the overlap region, the two methods give $v_2$
values that can be as much as 50\% apart for the 0-5\% central Au+Au
data. Thus, in order to determine background fluctuation we use the
average of the two values, and then use the maximum and minimum
values to estimate the systematic uncertainty in the yield
measurement. For the purposes of the study, we assume constituent
quark scaling \cite{bib6}, and use the $v_2$ measured for pions.
Thus the difference in the background between the species is due
only to the difference in the level of the uncorrelated background,
which will be discussed in the next section.

\subsection{Background subtraction: Setting the background
level}\label{sect:bgSubtr}

In each collision system considered here, there was a large
uncorrelated background associated with each correlation function.
The height of the background can either be measured by fitting the
correlation function, or by assuming Zero Yield At Minimum (ZYAM)
\cite{bib7}. The ZYAM method works under the assumption that at the
minimum $x_{min}$ of the correlation function, $C_n$(\df),
$C_n(x_{min})$ is due only to background. The determination of the
function's minimum and the differences between a free fit and a ZYAM
calculation, introduce an additional source of systematic
uncertainty on the same-side yield. In the \dau\ data the statistics
were scarce and thus the statistical error bars on the data points
were too large to constrain $x_{min}$ precisely.  Thus only a
periodic fit with an ansatz as shown in equation \ref{eq:corrFnDAu}
was used.

\begin{equation}
C_n(\Delta\phi)=A_0e^{-\Delta\phi^2/2\sigma}+A_0e^{-(\Delta\phi-\pi)^2/2\sigma}+B
\label{eq:corrFnDAu}
\end{equation}

    In \auau\ data the ansatz had an additional term: $B$ in equation \ref{eq:corrFnDAu} was replaced
    by $B\times (1+2v_2^{assoc}v_2^{trigg}\cos{2\Delta\phi})$.
    Using this ansatz, we calculate the $v_2$ coefficients associated with the given mean \pt\ of the
    trigger and associated particles.  We then estimate the minima of the correlation
    function by fitting two, three or four points around the trough.
     We then set our ansatz equation to zero and calculate $B$.  At this
    point we know the underlying term completely, and can subtract
    the background function point by point from the experimental
    correlation function.  We then fit the same-side of the result
    using a Gaussian, and calculate the yield.

\section{Results}\label{tabs}

\begin{figure}\centering 
                 \includegraphics[width=8cm]{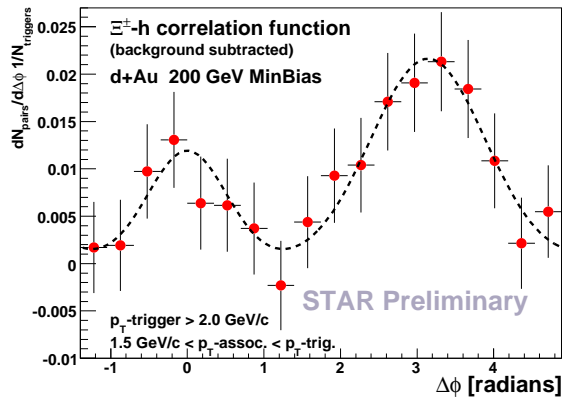}
 \caption[]{$\Xi^{\pm}$-h correlation function in
$\sqrt{s_{NN}}=$200 GeV $d$+Au data, normalized by the number of
triggers.  The function is shown after background subtraction.
 The dashed curve shows the two-gaussian free fit to the data, from which the yields were
 extracted. All $\Xi$ triggers are at \pt\ $>2$ GeV/c, all associated particles have 1.5
 GeV/c $<$ \pt\ $<$\pt\ -trigger.}
\label{fig:dAu} \vspace*{-0.6cm}
\end{figure}

Multi-strange baryon azimuthal correlations were observed in $d$+Au
and Au+Au data and same-side yields extracted for a variety of
trigger \pt.\  Figure \ref{fig:dAu} shows the correlation function
constructed using $\Xi$ baryons triggers in \dau\ data.  To gain
more statistics, the trigger \pt\ threshold was lowered to 2 GeV/c.
The statistics are still scarce, but after fitting the correlation
function as described in section \ref{sect:bgSubtr}, we were able to
extract the same side-yield per trigger, which was $0.015\pm0.026$.
This is approximately ten times smaller than the same-side yield
obtained at the same trigger and associated particle \pt\ in 0-10\%
central \auau\ data ($0.20\pm0.05$).

\begin{figure}\centering
  \includegraphics[width=11cm]{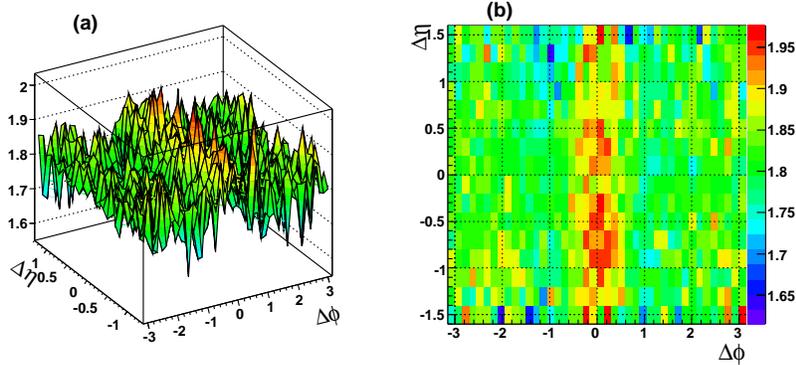}\\
  \vspace*{-0.2cm}
  \caption{\de-\df\ correlation with $\Xi$ triggers with 2.5 $<$ \pt\-trig $<$ 6.5 GeV/c,
  and associated particles with 2.0 $<$ \pt\-assoc. $<$ \pt\-trig. The two panels illustrate
  the existence of an elongation in \de\ on the same side of the azimuthal
  correlation (ridge), as well as the statistics, which do not allow for separation
  of the ridge from the jet signal.  In panel a) the plot is shown in three dimensions, panel b)
  shows the plot from above (color on-line).}\vspace*{-0.6cm}\label{fig:2DXi}
\end{figure}

Looking at the \auau\ $\Xi$ correlation function in two dimensions
(as shown in Figure \ref{fig:2DXi}), we see that the peak around
\df=0 is elongated in \de.\ This underlying structure, or ridge, has
been previously observed in correlations constructed using charged
particles and \La\ baryons. This is the first observation of this
effect for multi-strange triggered correlations. To separate any jet
peak from the underlying ridge, we divide the 2-dimensional
phase-space into a jet+ridge and ridge-only regions. We then
subtract the ridge-only space from that where both jet and ridge are
present. The method has been effectively used previously for \La\
and charged particles \cite{bib8}. However, in the case of
multi-strange baryons, the statistics are insufficient to extract
the jet yield from the jet+ridge combination. From measurements of
\La\ and $K^0_S$, \cite{bib8} we know that at similar \pt-trigger
and \pt-associated, the jet yield is approximately independent of
the size of the collision. Assuming this holds true for the
multi-strange particles, we can approximate the contribution of the
ridge to the yield by comparing the \dau\ yields to those obtained
at similar \pt\-trigger in central Au+Au. Comparing the two yields,
we observe that at least 90\% of the same-side $\Xi$ correlation
peak in central \auau\ comes from the ridge.

\begin{figure}\centering
  \includegraphics[width=9cm]{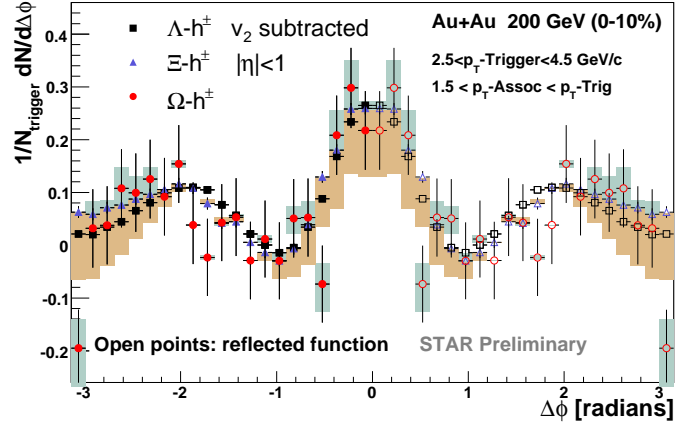}\\
 \vspace*{-0.2cm}
  \caption{\df\ correlation functions with \La,\ $\Xi$, and \Om\ triggers at 2.5 $<$ \pt\-trig $<$ 4.5 GeV/c,
  and associated particles with 1.5 $<$ \pt\-assoc. $<$ \pt\-trig. The blue and brown bands show the
  systematical error due to uncertainty in determining the $v_2$ coefficient. }\vspace*{-0.6cm}\label{fig:SubtrCorr}
\end{figure}

 Given the prediction in \cite{bib1}, we would expect
that the same-side peak observed in $\Xi$ correlations is due to the
non-strange quark present in the $\Xi$ baryon.  We therefore look at
the triply-strange \Om\ baryons to test the prediction.  Following
the same analysis technique as before, we obtain a result shown in
Figure \ref{fig:SubtrCorr}. The figure presents \La\, $\Xi$, and
\Om\ azimuthal correlations, with systematic uncertainties due to
flow determination methods.  For clarity, the uncertainty is plotted
for $\Xi$ and \Om\, but not for \La.\ As the correlation functions
are symmetric around 0 and $\pi$ radians, we double our statistics
by ``folding" the functions around 0. Thus, the ``reflected" side is
shown using open symbols, while the ``folded" side is shown in
solid.

 Figure \ref{fig:SubtrCorr} clearly shows the
same-side peak for \Om\ baryon-triggered correlation.  Moreover,
within errors, there are no observable differences in peak sizes
between the three species.

\begin{figure}\centering
  \includegraphics[width=10cm]{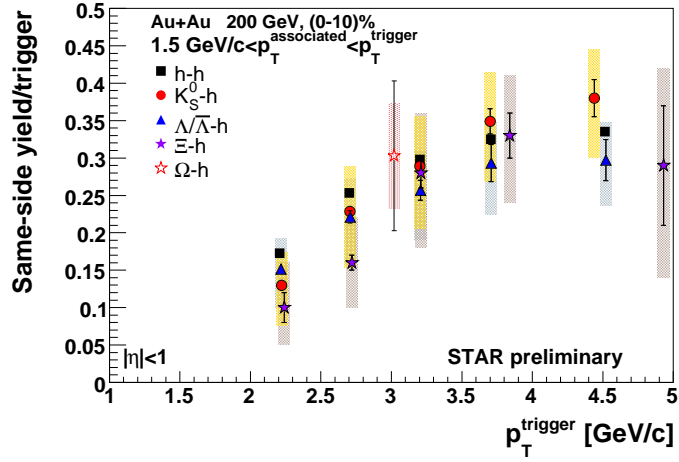}\\
 \vspace*{-0.5cm}
  \caption{Jet+Ridge yields of strange particles in 0-10\% central Au+Au collisions.
  The shaded bands represent the systematic error due to the method used in determining the
  elliptic flow coefficient (color on-line).  Yellow bands correspond to the error on the
  $K^0_S$ measurement, light blue -- the error on the \La\ measurement.  The brown bands
  represent the error on the $\Xi$ determination, and pink -- on that of the \Om\ baryon.   }\label{fig:All Yields}
\end{figure}
We study the trigger \pt\ and species dependence where there are
sufficient statistics.  Figure \ref{fig:All Yields} shows
unidentified charged particle, $K^0_S$, \La,\ $\Xi$, and \Om\
same-side yields per trigger, as a function of trigger \pt.\ The
systematic uncertainty due to $v_2$ determination methods is
represented by shaded bands.  The yields appear to rise
systematically with increasing \pt,\ however, there is no observable
species dependence.  Indeed, the \Om\ yield is consistent with the
same-side yields of all the other particles, including unidentified
charged hadrons.  For the latter, we associate this behavior with
jet production. Moreover, charged particle correlations are also
accompanied by a similar broadening of the signal in the delta-eta
space we observe for the multi-strange baryons.  Thus, we conclude
that if the multi-strange particles observed in most central Au+Au
collisions at intermediate \pt\ are not made in jets, they are at
least associated with jet production.

\section{Conclusions and Outlook}\label{figs} In this work, we have
presented the first measurement of multi-strange baryon azimuthal
and longitudinal correlations in relativistic heavy-ion collisions.
Correlation functions were obtained in the $d$+Au and the central
0-10\% Au+Au data sets. The yields measured for $\Xi$ baryon
correlations in the $d$+Au data are consistent with other strange
particle results in the same data and with jet yields of
singly-strange particles in the same trigger-momentum region of the
most central Au+Au data. However, the triply-strange baryon
correlation is the main result presented in this work. An $\Omega$
baryon correlation function was obtained in the 0-10\% most central
Au+Au collisions and after background subtraction, the same-side
peak was compared to that of the $\Xi$ baryon correlation and
$\Lambda$ baryon correlation in the same data.  Contrary to
predictions \cite{bib1}, the $\Omega$ same-side peak yield is
$\sim2\sigma$ above a null value. In fact, the per-trigger yields of
the three strange baryon species, \La\, $\Xi$, and $\Om$ are
consistent within errors.  However, there are caveats. Although
there are sufficient statistics to {\it observe} a $\Xi$ baryon
correlation function in $\Delta\eta-\Delta\phi$ space in the most
central Au+Au collisions, we are not yet ready to draw definitive
conclusions as to whether there is a jet-only component of the
multi-strange correlation functions.  New calculations, introduced
since the \Om\ correlation was first presented \cite{bib8}, predict
that the same-side \Om\ peak is due to the jet interaction with the
medium, the so-called ``phantom jet" \cite{bib9}, not to a jet where
\Om\ baryon is the leading particle. Further studies of
multi-strange correlations in $\Delta\eta$ space will yield more
details, however at present the available statistics do not allow
for such measurement.

\vfill\eject
\end{document}